\documentclass[aps,floatfix,twocolumn,showpacs]{revtex4}
\usepackage{graphicx,amssymb,amsmath}
\newcommand{\leftsub}[2]{{\vphantom{#2}}_{#1}{#2}}

\begin{document}

\title{Distinguishability of Hyper-Entangled Bell States by Linear Evolution and Local Projective Measurement}
\author{N. Pisenti}
\author{C.P.E. Gaebler}
\author{T.W. Lynn}
\email{lynn@hmc.edu}
\affiliation{Department of Physics, Harvey Mudd College, 301 Platt Blvd., Claremont, California 91711, USA}
\date{\today }

\begin{abstract}
Measuring an entangled state of two particles is crucial to many quantum communication protocols. Yet Bell state distinguishability using a finite apparatus obeying linear evolution and local measurement is theoretically limited. We extend known bounds for Bell-state distinguishability in one and two variables to the general case of entanglement in $n$ two-state variables. We show that at most $2^{n+1}-1$ classes out of $4^n$ hyper-Bell states can be distinguished with one copy of the input state. With two copies, complete distinguishability is possible.  We present optimal schemes in each case.
\end{abstract}
\maketitle

% ----------------------------------------------------------------
 
\section{Introduction}
\label{Section:Intro}
Entangled systems are ubiquitous in quantum information science, playing key roles in teleportation~\cite{TeleportationProtocol}, quantum repeaters~\cite{QuantumRepeaters}, dense coding~\cite{DenseCoding}, entanglement swapping~\cite{entswapping,entswappingexpt}, and fault tolerant quantum computing~\cite{QuantumComputing}. Typical barriers to efficiently realizing these applications are twofold---first, the reliable generation of entangled pairs in a particular Bell state, and second, complete Bell-state measurement between  two particles \cite{BellMeasurements,MethodsTeleportation}.  Entangled pair creation can be achieved via numerous methods; for example, with photons it is possible through the non-linear interactions involved in spontaneous parametric downconversion \cite{KwiatSPDC}. However, a complete, deterministic Bell-state measurement is impossible within the broad class of apparatus obeying linear evolution and local measurement (LELM)~\cite{BellMeasurements,MethodsTeleportation}. Much focus is placed on these devices nonetheless, due to their ease of implementation.  The inability to perform a complete, deterministic Bell-state measurement with LELM has limited the unconditional fidelity achieved in numerous experimental settings \cite{entswappingexpt,densecodingexpt,superdenseexpt,tpexpt}. A deeper understanding of the exact bounds placed on Bell-state distinguishability by LELM devices thus has implications for quantum communication protocols and other applications in quantum information science. 

Recent experimental developments have opened the arena of entanglement between two particles in multiple degrees of freedom, a circumstance known as hyper-entanglement \cite{hyperE}. Existing bounds on nonlocal state distinguishability have involved systems entangled in two or fewer two-state variables \cite{BellMeasurements,MethodsTeleportation,HE}, or in one three-state or $n$-state variable \cite{Calsa-general,vanLoockLut,Carollo1,Carollo2}, yet experiments to date have achieved entanglement in up to three variables~\cite{hyperEexpt}. Thus there is considerable motivation for more general theoretical bounds, for instance to establish channel capacities for superdense coding. In this paper, we consider the general case of two particles entangled in $n$ two-state variables. Our analysis offers an $n$-variable distinguishability limit based on a simple understanding of the restrictions imposed by LELM; we further describe a straightforward apparatus which will always achieve maximum distinguishability between hyper-entangled Bell states.

\section{Notation and representation of LELM apparatus}

An apparatus constrained by ``linear evolution'' acts on each input particle independently of the other, so it can be represented as a unitary transformation over the space of single-particle input states.  Consequently, the single-particle output modes of the device are linear combinations of the single-particle input modes. ``Local projective measurement''  means the detection event projects the system into a product state of two single-particle output modes.  We consider measurement in a Fock state basis of output modes, corresponding to annihilation of particles in the two 
detectors which register clicks.

The system of interest consists of two particles whose states are described by $n$ two-state variables, each of which is represented in the basis $\{|0\rangle, |1\rangle\}$. Some examples include the following: for photonic systems, the linear polarization states $\{H, V\}$, the subset $\{ +\hbar, -\hbar\}$ of orbital angular momentum states, or time bins $\{t_s, t_l\}$; for atomic systems, two ground or metastable electronic states $\{g_1,g_2\}$; for electronic spin qubits, the states $\{\uparrow_z,\downarrow_z\}$; and many more two-state quantum systems.  

\begin{figure}[t]
	\begin{center}
		\includegraphics[width=2.75in, trim = 0in 0in 0in 0in, clip=true]{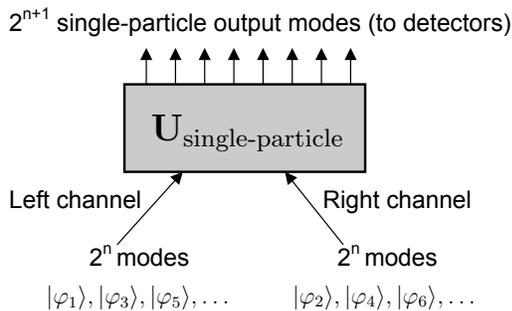}
		\caption{A pair of particles enters the measurement apparatus via separate channels (Left and Right).  Each particle evolves independently of the other (linear evolution), hence the unitary evolution of single-particle input modes to output modes.   Local measurement registers two clicks in the detectors, projecting the system into a product state of two single-particle output modes (possibly the same mode twice).
		\label{schematic}}
	\end{center}
\end{figure}

The two particles enter the LELM apparatus via separate spatial channels, designated \textit{L} and \textit{R}, as shown schematically in Fig.~\ref{schematic}.  Each single-particle input undergoes unitary evolution to the set of orthogonal output modes.   Each detector is capable of resolving number states in its associated mode, so two particles in a single detector can be reliably detected. A complete single-particle input is specified by particular values for all $n$ variables as well as the spatial channel.  It has been shown that, for projective measurements with linear evolution, distinguishability between signal states cannot be improved by the use of auxiliary modes as long as the signal states are of definite particle number \cite{vanLoockLut,Carollo2}.  Thus we may restrict our discussion to a space of single-particle input modes with dimension $2^{n+1}$, and a corresponding $2^{n+1}$-dimensional space of output modes. Finally, a complete detection event consists of annihilating particles in two output modes (possibly the same mode twice).

A useful basis for the input states consists of kets $|\varphi_m\rangle$, each representing a particle in one of the $2^{n+1}$ possible input modes:  either the $|0\rangle$ or $|1\rangle$ eigenstate of each variable, and either the left or right input channel.  We assign odd indices $m$ to \textit{L}-channel states and even indices to \textit{R}-channel states, such that $|\varphi_{2s-1}\rangle = |\chi_s,L\rangle$ and $|\varphi_{2s}\rangle = |\chi_s,R\rangle$ are identical to one another except for the choice of left vs. right input channel; $s$ ranges from $1$ to $2^n$ and $\{\chi_s\}$ is the set of all binary strings of length $n$.  For example, for $n=1$ (one variable), the input-state basis is:
\begin{equation}
|\varphi_1\rangle = |0,L\rangle, ~~
|\varphi_2\rangle = |0,R\rangle, ~~
|\varphi_3\rangle = |1,L\rangle, ~~
|\varphi_4\rangle = |1,R\rangle. \label{eq:1dinputbasis}
\end{equation}

The two-particle, or overall, input states are spanned by the set of tensor product states $|\varphi_m\rangle |\varphi_k\rangle$ with the restriction that $m \neq k~(\mathrm{mod}~2)$, since the input states of interest are limited to those with one particle in each of the left and right input channels.  For indistinguishable particles 1 and 2, the \mbox{(anti)symmetrized} version $\frac{1}{\sqrt{2}}(|\varphi_m\rangle_1|\varphi_k\rangle_2 \pm |\varphi_k\rangle_1|\varphi_m\rangle_2)$ is understood instead.

We can describe a click in detector $i$ as a projection of the input state onto the single-particle output mode $|i\rangle$. The relationship between input and output modes depends on the apparatus, but without loss of generality we can write 
\begin{equation}
	|i\rangle = \sum_{m} U_{im}|\varphi_m\rangle \label{U-transform}
\end{equation}
where the LELM apparatus is represented by the unitary matrix $\mathbf{U}$.  Each output mode thus takes the form 
\begin{equation}
|i\rangle = \alpha_i|l_i\rangle + \beta_i|r_i\rangle, 
\label{det-even-odd}
\end{equation}
where $|l_i\rangle$ is a superposition of left-channel input states and $|r_i\rangle$ is a superposition of right-channel input states.

A complete detection signature corresponds to a projection of the two-particle input state onto the tensor product state $|i\rangle|j\rangle$; for indistinguishable particles, the \mbox{(anti)symmetrized} version $\frac{1}{\sqrt{2}}(|i\rangle_1|j\rangle_2 \pm |j\rangle_1|i\rangle_2)$ is understood instead.  Furthermore, if we constrain the inputs to include just one particle in the left channel and one particle in the right, we should consider the projection of $|i\rangle|j\rangle$ onto the subspace of two-particle input states spanned by $|\varphi_m\rangle|\varphi_k\rangle$ with $m\neq k~(\mathrm{mod}~2)$.  We call this projection the detection signature, denoted $P_{LR}|i\rangle|j\rangle$.

Finally, we turn to a description of the Bell states themselves, which form an entangled basis for the two-particle system.  In a single variable, the Bell basis consists of four maximally-entangled states given by 
\begin{align}
	 |\Phi^{\pm}\rangle &=\frac{1}{\sqrt{2}}\Big(|0,L\rangle|0,R\rangle\pm |1,L\rangle|1,R\rangle\Big) \label{Equation:PhiBell}\\
	 |\Psi^{\pm}\rangle &=\frac{1}{\sqrt{2}}\Big(|0,L\rangle|1,R\rangle \pm |1,L\rangle|0,R\rangle\Big), \label{Equation:PsiBell}
\end{align}
or rather the (anti)symmetrized versions of Eqs.~\ref{Equation:PhiBell}~and~\ref{Equation:PsiBell}. To generalize this basis for hyper-entanglement in $n$ variables, we simply take a tensor product between the Bell states for each individual variable. Thus, the hyper-Bell states are $\{|\Phi^+\rangle,|\Phi^-\rangle,|\Psi^+\rangle,|\Psi^-\rangle\}^{\otimes n}$; for $n$ variables, these are $4^n$ mutually orthogonal entangled states.

\section{MAXIMUM NUMBER OF DISTINGUISHABLE BELL-STATE CLASSES}
 
As discussed above, there are $2^{n+1}$ mutually orthogonal output modes, or $2^{n+1}$ detectors.  Each hyper-Bell state, due to its maximal entanglement, is capable of producing at least one click in any detector $i$. To see this, suppose that detector $i$ is never triggered by  hyper-Bell state $|B\rangle$.  Then 
$\langle B|\frac{1}{\sqrt{2}}(|j\rangle_1|i\rangle_2 \pm |i\rangle_1|j\rangle_2)=0$ for all output modes $|j\rangle$, or equivalently, 
\begin{equation}
\langle B_{\text{sym}}|\big(|j\rangle_1|i\rangle_2\big)=0  ~~ \forall j
\label{eq:todisprove}
\end{equation} 
where $|B_{\text{sym}}\rangle$ is the Bell state symmetrized or antisymmetrized under exchange of particles 1 and 2.  For example, the symmetrized version of the single-variable Bell state $|\Phi^+\rangle$ is 
\begin{align}
|\Phi^+_{\text{sym}}\rangle = \frac{1}{2} \Big( &|0,L\rangle_1 |0,R\rangle_2 + |1,L\rangle_1|1,R\rangle_2 \notag
	\\ &+ |0,R\rangle_1|0,L\rangle_2 + |1,R\rangle_1|1,L\rangle_2\Big).
	\label{eq:phiplussym}
\end{align}

If Eq. \ref{eq:todisprove} holds, it follows that 
\begin{equation}
	\sum_{j} \big( \leftsub{2}{\langle i|} \leftsub{1}{\langle j|} \big) |B_{\text{sym}}\rangle \langle B_{\text{sym}}|
	\big( |j\rangle_1|i\rangle_2\big)=0.
\end{equation}
However, the left-hand side of the last expression is simply $_2\langle i|Tr_1(|B_{\text{sym}}\rangle\langle B_{\text{sym}}|)|i\rangle_2$, where the trace is taken over the states of particle 1.  However, this quantity cannot be zero since the reduced density matrix $Tr_1(|B_{\text{sym}}\rangle\langle B_{\text{sym}}|)$ is a multiple of the identity on the space of particle~2 states, including both left- and right-channel states.  (Consider, for example, $Tr_1(|\Phi^+_{\text{sym}}\rangle\langle \Phi^+_{\text{sym}}|)$ calculated using Eq. \ref{eq:phiplussym}.) Thus our supposition fails: Eq. \ref{eq:todisprove} cannot hold, and so every Bell state can in fact trigger every detector.  A single detector click cannot discriminate between any of the Bell states.

An alternate demonstration of this key point proceeds as follows.  If the initial two-particle state is an arbitrary Bell state $|B\rangle$, the state following a single click in some detector is proportional to $\hat{c}|B\rangle$, where $\hat{c}$ is the annihilation operator associated with that output mode.  The statement that the Bell state can cause the detector to click is equivalent to the statement that the norm of this post-click state is nonzero.  Thus we must consider the quantity $\langle B|\hat{c}^{\dagger} \hat{c}|B\rangle$.  We can rewrite the expression in terms of the annihilation operators $\hat{a}_m$ associated with the single-particle input modes $|\varphi_m\rangle$:  $\hat{c} = \sum_{m} C_m \hat{a}_m$.  (Recall from Sec. II that $m$ odd or even denotes left- or right-channel modes, respectively.  Further, if we are considering the output mode associated with detector $i$, then $C_m = U^{\ast}_{im}$ in the notation of Eq. \ref{U-transform}.) Thus
\begin{align}
 \langle B|\hat{c}^{\dagger} \hat{c}|B\rangle = &\sum_{m,k} C^{\ast}_m C_k \langle B|\hat{a}^{\dagger}_m \hat{a}_k|B\rangle \notag \\
 = &\sum_{m} |C_m|^2 \langle B|\hat{a}^{\dagger}_m\hat{a}_m|B\rangle \notag \\
 &+\sum_{m,k\neq m} C^{\ast}_m C_k \langle B|\hat{a}^{\dagger}_m \hat{a}_k|B\rangle.  
\label{eq:postnorm}
\end{align}

To further evaluate this expression, we write the Bell state as 
\begin{equation}
|B\rangle = \frac{1}{\sqrt{2^n}} \sum^{2^n}_{s=1} (-1)^{\sigma_B(s)}\hat{a}^{\dagger}_{2s-1} \hat{a}^{\dagger}_{2r_B(s)} |\mathbf{0}\rangle
\label{eq:Bellfromvac}
\end{equation}
where $|\mathbf{0}\rangle$ is the vacuum state, $\sigma_B(s)$ can take values 0 or 1, and $\{r_B(s)\}$ is a permutation of $\{s\}$, so each left-channel mode of index $2s-1$ is uniquely paired in this Bell state with a right-channel mode of index $2r_B(s)$ (and vice versa).  From this form it is easy to see that, for any $k\neq m$, $\hat{a}_k|B\rangle$ and $\hat{a}_m|B\rangle$ are orthogonal to each other, so $\langle B| \hat{a}^{\dagger}_m \hat{a}_k |B\rangle=0$.  The final sum in Eq. \ref{eq:postnorm} vanishes by this reasoning.  In the remaining sum of Eq. \ref{eq:postnorm}, $\langle B| \hat{a}^{\dagger}_m\hat{a}_m|B\rangle = \frac{1}{2^n}$ for any input mode $m$ and any Bell state $|B\rangle$, and so the sum simply evaluates to $\frac{1}{2^n}\sum_m |C_m|^2 = \frac{1}{2^n}$.  Thus in the end we have $\langle B|\hat{c}^{\dagger} \hat{c} |B\rangle = \frac{1}{2^n}$, a nonzero value independent of the particular Bell state and output mode.  In particular, any output mode is compatible with all Bell states:  a single detector click does not discriminate between Bell states.   

\vskip1em

Because a single detector event provides no information about which hyper-Bell state the particles occupy, distinguishability must come from identifying one of the $2^{n+1}$ orthogonal outcomes for the second detector event. The $2^{n+1}$ possibilities form a simple upper bound on distinguishable Bell-state classes from LELM devices, obtainable also by considering the Schmidt number of at most 2 for any detection signature \cite{Calsa-general}.  We will now show that the actual maximum is one less than the simple upper bound, namely, $2^{n+1}-1$. This general result agrees with previous results for $n=1$ (3 out of 4) and $n=2$ (7 out of 16)  \cite{MethodsTeleportation,BellMeasurements,HE}.

For fermions, the amplitude to observe two clicks in detector $i$ must always be zero, since $|i\rangle|i\rangle$ is inherently symmetric under particle exchange.  Thus for any detector $i$, at most $2^{n+1}-1$ detection signatures $P_{LR}|i\rangle|j\rangle$ are nonzero.  Since all the Bell states are represented in these $2^{n+1}-1$ signatures, there are at most $2^{n+1}-1$ distinguishable classes of hyper-Bell states for two fermions.

For bosons, consider a single output mode $|i\rangle$ as represented in Eq. \ref{det-even-odd}.  If either coefficient $\alpha_i$ or $\beta_i$ is zero, the detection signature $P_{LR}|i\rangle|i\rangle$ is zero, and we have at most $2^{n+1}-1$ distinguishable classes of Bell states.

If $|i\rangle$ is a nontrivial superposition of left- and right-channel inputs as in Eq. \ref{det-even-odd}, then some linear combination of output modes must satisfy:
\begin{equation}
|X\rangle = \sum_{j} \epsilon_j|j\rangle = \alpha_i|l_i\rangle - \beta_i|r_i\rangle.
\label{otherstate}
\end{equation}
The hypothetical detection signature $P_{LR}|i\rangle|X\rangle$ is zero, giving $\sum_{j} \epsilon_j P_{LR}|i\rangle|j\rangle = 0$.  Consider some $j$ such that $\epsilon_j \neq 0$; any Bell state represented in the detection signature $P_{LR}|i\rangle|j\rangle$ must also be represented in at least one other detection signature $P_{LR}|i\rangle|k\rangle$.  Thus it is not possible to reliably distinguish between a class of Bell states that can produce clicks in detectors $(i,j)$ and a class that can produce clicks in detectors $(i,k)$.  Therefore the number of distinguishable Bell-state classes must be less than the full number of detection signatures involving detector $i$, and there can be at most $2^{n+1}-1$ distinguishable classes of hyper-Bell states for two bosons.

If the left and right input channels are not brought together in the apparatus, \textit{e.g.}, for experimental convenience, each output mode $|i\rangle$  of Eq. \ref{det-even-odd} is simply equal to $|l_i\rangle$ or to $|r_i\rangle$. 
Thus $2^n$ output modes are superpositions of the left-channel inputs, and the other $2^n$ output modes are superpositions of the right-channel inputs.  For any detector $i$, only $2^n$ detection signatures $P_{LR}|i\rangle|j\rangle$ are nonzero.  Since all the Bell states are represented in these $2^n$ signatures, there are at most $2^n$ distinguishable classes of Bell states for this case.  

\section{APPARATUS FOR MAXIMAL DISTINGUISHABILITY}

A best-case apparatus for separate \textit{L} and \textit{R} measurement can be achieved by measuring the \textit{L}- and \textit{R}-channel inputs each in the $\{|0\rangle,|1\rangle\}^{\otimes n}$ basis.  This is a projective measurement in the $|\varphi_m\rangle$ basis, so detection signatures are of the form $|\chi_s,L\rangle|\chi_t,R\rangle$.  Bell states represented in one detection signature must be tensor products of $|\Phi^\pm\rangle$ in variables where $|\chi_s\rangle$ and $|\chi_t\rangle$ share their eigenvalue, and $|\Psi^\pm\rangle$ in variables where they do not.  Specifying $\Phi$ vs. $\Psi$ in this way yields $2^n$ classes of $2^n$ Bell states each.

A unitary transformation realizing the maximal \mbox{$2^{n+1}-1$} Bell-state classes for fermionic or bosonic inputs is given by ($s=1$ to $2^n$):
\begin{align}
|2s-1\rangle &= \frac{1}{\sqrt{2}}(|\varphi_{2s-1}\rangle + |\varphi_{2s}\rangle) =\frac{1}{\sqrt{2}}(|\chi_s,L\rangle + |\chi_s,R\rangle) \notag \\
|2s\rangle &= \frac{1}{\sqrt{2}}(|\varphi_{2s-1}\rangle - |\varphi_{2s}\rangle)=\frac{1}{\sqrt{2}}(|\chi_s,L\rangle - |\chi_s,R\rangle).
\label{eq:besttransform}
\end{align}
This is a Hadamard transform between the \textit{L} and \textit{R} channels for each $n$-variable eigenstate $|\chi_s\rangle$. 

\begin{figure}[t]
	\begin{center}
		\includegraphics[width=2.6in, trim = 0in 0in 0in 0in, clip=true]{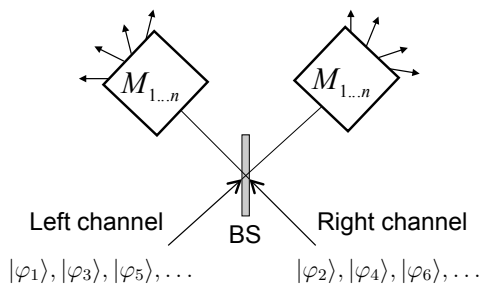}
		\caption{Apparatus for optimal hyper-Bell state distinguishability of photon pairs as in Eq. \ref{eq:besttransform}.  The 50/50 beamsplitter interferes $|\varphi_{2s-1}\rangle$ and $|\varphi_{2s}\rangle$.  Evolution $M_{1\ldots n}$ separates particles by values of the $n$ variables, so detection is a projective measurement in the $\{|0\rangle,|1\rangle\}^{\otimes n}$ basis.  (If the beamsplitter does not preserve variables ${1\ldots k}$, $M_{1\ldots k}$ must precede the beamsplitter stage while $M_{k+1\ldots n}$ can follow.)  For one-variable Bell states of photon polarization, $M_{1\ldots n}$ is simply $M_{1}$ and can be realized with a polarizing beamsplitter.      
		\label{fig:optimaldet}}
	\end{center}
\end{figure}

For bosons with linear evolution governed by Eq. \ref{eq:besttransform}, a detection signature $P_{LR}|2s-1\rangle|2s-1\rangle$ or $P_{LR}|2s\rangle|2s\rangle$ identifies the two-particle input state $|\chi_s,L\rangle|\chi_s,R\rangle$.  These detection signatures thus all identify the class of $2^n$ hyper-entangled Bell states $|\Phi^\pm\rangle^{\otimes n}$.  Detection signatures of the form $P_{LR}|2s-1\rangle|2s\rangle$ or $P_{LR}|2s\rangle|2s-1\rangle$, however, are antisymmetric under particle exchange and do not occur.  For fermions the roles are reversed; detection signatures $P_{LR}|2s-1\rangle|2s\rangle$ or $P_{LR}|2s\rangle|2s-1\rangle$ identify the class of $2^n$ hyper-entangled Bell states $|\Phi^\pm\rangle^{\otimes n}$, while detection signatures $P_{LR}|2s-1\rangle|2s-1\rangle$ or $P_{LR}|2s\rangle|2s\rangle$ are symmetric and do not occur.

Any detection signature not of the forms already discussed will give
\begin{equation}
P_{LR}|i\rangle|j\rangle  = \frac{1}{\sqrt{2}}(|\chi_s,L\rangle|\chi_t,R\rangle \pm |\chi_t,L\rangle|\chi_s,R\rangle)
\label{smallclasses}
\end{equation}
 with $t\neq s$. Bell states represented in such a detection signature have a well-defined sequence of $\Phi$ vs. $\Psi$ in the $n$ variables.  Furthermore, the sign of the superposition in Eq. \ref{smallclasses} gives the symmetry of the overall state with respect to exchange of \textit{L} and \textit{R}; the $|\Psi^-\rangle$ Bell state is antisymmetric in this way while the others are all symmetric, so a $+$ ($-$) sign in Eq. \ref{smallclasses} restricts that detection signature to hyper-entangled Bell states with $|\Psi^-\rangle$ in an even (odd) number of variables.  Thus each such detection signature identifies a class of $2^{n-1}$ Bell states.  There are $2^{n+1}-2$ classes of this type and one $|\Phi^\pm\rangle^{\otimes n}$ class, so exactly $2^{n+1}-1$ classes are reliably distinguished.

An optimal apparatus for photons is depicted in Fig.~\ref{fig:optimaldet}. A 50/50 beamsplitter performs (up to overall phase shifts) the \textit{L}/\textit{R} Hadamard transform of Eq. \ref{eq:besttransform}; the input modes are then separated according to the value of each variable, so detector clicks project into the $\{|0\rangle,|1\rangle\}^{\otimes n}$ basis.  Previous optimal distinguishability schemes for $n=1$ are of this form \cite{Innsbruck1,Innsbruck2,Innsbruck3,MethodsTeleportation,BellMeasurements}.

The schemes above can be varied by performing projective measurement in the diagonal $\{\frac{1}{\sqrt{2}}\left(|0\rangle + |1\rangle\right),\frac{1}{\sqrt{2}}\left(|0\rangle - |1\rangle\right)\}$ basis rather than the $\{|0\rangle,|1\rangle\}$ basis for one or more variables. For example, for $n=1$, an optimal unitary transformation can be written in the basis of Eq. \ref{eq:1dinputbasis} as
\begin{equation}
	\label{Equation:UMatrix2}
	\mathbf{U_{opt}}=\frac{1}{2}\begin{pmatrix}1 & 1 & 1 & 1 \\ 1 & -1  & 1 & -1 \\ 1 & 1 & -1 & -1 \\ 1 & -1 & -1 & 1 \end{pmatrix}.
\end{equation}
For polarization-entangled photons, Eq. \ref{Equation:UMatrix2} is realized by a 50/50 beamsplitter and polarizing beamsplitters at $45^{\circ}$. 

With two copies of a hyper-Bell state, complete distinguishability can be achieved by measuring one copy in the apparatus of Eq. \ref{eq:besttransform}, and the second in a version where detectors project into the diagonal basis for all $n$ variables.  $|\Phi^+\rangle$ and $|\Psi^-\rangle$ in each variable retain their forms in the diagonal basis, while the other Bell states exchange forms, $|\Phi^-\rangle \leftrightarrow |\Psi^+\rangle$.  States indistinguishable via the first apparatus share a common sequence of $\Phi$ vs. $\Psi$ in the $n$ variables, and differ by $+$ vs. $-$ in one or more variables.  All these states will have distinct $\Phi$ vs. $\Psi$ sequences in the diagonal basis, so they will give distinct measurement outcomes in the second apparatus. In fact, the same principle gives two-copy complete distinguishability even by measuring L- and R-channel inputs separately in the $\{|0\rangle,|1\rangle\}^{\otimes n}$ basis for one copy and the $\{\frac{1}{\sqrt{2}}\left(|0\rangle + |1\rangle\right),\frac{1}{\sqrt{2}}\left(|0\rangle - |1\rangle\right)\}^{\otimes n}$ basis for the second copy.  Other optimal schemes exist, such as those already known for $n=2$ \cite{KW,KWexpt,HE}.   

\section{Conclusion}
We have shown that devices constrained by linear evolution and local measurement cannot reliably distinguish more than $2^{n+1}-1$ Bell states or Bell-state classes for two particles entangled in $n$ degrees of freedom.  This bound holds even for conditional measurements; after a click in detector $i$, no conditional evolution of the remaining state to output channels will avoid the limitations presented above.  However, two copies of the hyper-Bell state allow complete distinguishability.  We have constructed unitary transformations of input states to output states which achieve the upper bound on distinguishability for one and two copies of the hyper-Bell state.

This work illustrates the potential and limitations for manipulation and measurement of entangled systems with inherently linear, unentangling devices.  It relies on a very physical approach to consider cases in which previous $n=1,2$ methods are computationally unattractive; checking the $n=3$ bound by previous methods involves searching for solutions to ${64\choose16}\approx 4.9\times10^{14}$ systems of 16 equations.  Our approach gives another way of understanding the $n=1,2$ bounds, and may provide a framework for further bounds, perhaps on LELM distinguishability of hyper-entangled states involving qutrit or higher-dimensional variables.

\section{Acknowledgments}

The authors thank M. Orrison and D. Skjorshammer for useful conversations, and an anonymous referee for thoughtful feedback and suggestions.  This work was supported by Research Corporation Cottrell College Science Grant No. 10598.

%\bibliography{BibPub}

\bibliographystyle{apsrev}

\end{document}